# Beyond lead halide perovskites: visible light photovoltaics with phase engineered bismuth-based oxide double-perovskites, $Bi_2MCrO_6$ (M = Fe, Mn)

N P Vikas[1], Ranjit K Pradhan[2,3], Somdutta Mukherjee[#], Udai P Singh[4], Biplab K Patra[2,3], Ravi P Srivastava[5], Amritendu Roy[1,*]

[1] School of Minerals, Metallurgical, and Materials Engineering, IIT Bhubaneswar, Jatni, Khurda-752050, Odisha, India

[2] Materials Chemistry Department CSIR-Institute of Minerals and Materials Technology Bhubaneswar, 751013, Odisha, India

[3] Academy of Scientific and Innovative Research (AcSIR), Ghaziabad, 201002, India

[4] School of Electronics Engineering, KIIT University, Bhubaneswar, 751024, Odisha, India

[5] Department of Materials Engineering, IIT Jodhpur, Rajasthan, 342030, India

[#] Independent Researcher

[*]Corresponding author: amritendu@iitbbs.ac.in

**Abstract**

Lead poisoning and notorious ambient instability in lead-based halide perovskites pave the way for the exploration of alternative materials for affordable and efficient solar cell fabrication. An important prerequisite to this end is the optoelectronic evaluation of the proposed material. Here we report, optoelectronic characterization of $Bi_2FeCrO_6$ (BFCO) and $Bi_2MnCrO_6$ (BMCO) thin films vis-à-vis performance of photovoltaic cells. Solution-deposited thin films (350-450 nm) of the above compositions demonstrate a double-perovskite structure with monoclinic $P2_1/c$ symmetry, albeit with mixed cation valences and deep-level defects. A thorough optoelectronic evaluation exhibits large optical absorption in the visible range ($\alpha \sim 10^4$-$10^5$ cm$^{-1}$), and high carrier density, ~$10^{17-20}$ cm$^{-3}$. Ultraviolet photoelectron spectroscopy measurement allowed determination of the positions of the band-edges (valence band maximum and conduction band minimum), required for the selection of carrier transport layers. In its first, BMCO-based FTO/$SnO_2$/BMCO/Spiro-OMeTAD/Ag solar cell produced a maximum 3.56% conversion efficiency. Using numerical simulation, we predict that with suitable defect control, the above conversion efficiency can increase significantly.

## 1. Introduction

Large photovoltaic response at an affordable price, along with minimal environmental impact, drive the materials research for next-generation solar photovoltaics (PV). While halide perovskites qualify in the former, lead poisoning and notorious ambient instability are the biggest impediments to their commercial viability. This justifies the search for novel PV materials and technologies.

Transition metal oxide (TMO) (double-)perovskites are exciting materials classes with attractive attributes such as chemical stability [1], abundance [2], reduced toxicity [3], large absorption coefficient [4], and versatility of the fabrication techniques [5,6]. These attributes make TMOs attractive for absorber layer in perovskite solar cells. In addition, appropriate selection of carrier transport layer materials is crucial for efficient charge collection and therefore, device performance [7]. Therefore, detailed investigation into the optoelectronic properties is an essential prerequisite for considering narrow-bandgap TMO double-perovskites as absorber layer in solar PV.

Seeing the role of lone $6s^2$ electron pair which contribute to the large dielectric constant, low carrier effective mass and defect tolerance in lead-based halide perovskites, iso-electronic bismuth-based oxide double-perovskites could be interesting test systems to explore efficient PV response [8]. A recent study by Sradhasgar et al. [9], predicted a large number of bismuth-based, narrow bandgap double-perovskite systems with prospective PV response. One such system, $Bi_2FeCrO_6$ (BFCO) has been explored for its prospective application in thin film solar cells [9–11]. A significant PV response in BFCO was first reported with monochromatic incident light (635 nm) exhibiting an open-circuit voltage ($V_{OC}$), ~0.6 V [12]. BFCO assumes a double-perovskite structure with site-disorder between similar sized $Fe^{3+}$ and $Cr^{3+}$ ions lowering its crystal symmetry [13]. Low-symmetry polar structure renders the likelihood of stabilization of ferroelectric (FE) phase and therefore opens up the possibility of its application in ferroelectric-photovoltaic (FE-PV) devices. Unlike traditional semiconductors, FE-based PV devices could achieve $V_{OC}$ values beyond their bandgaps attributed to the polarization-induced electric field that separates the electron-hole pairs [14]. The polarization-induced electric field, further controls the chemical potential and surface band bending, thereby tailor the conversion efficiency [13]. Li et al. showed that the electronic structure of BFCO consists of Cr3d –O2p hybridized states forming the valence band (VB) and the empty Fe-3d making the conduction band (CB). Tuning the bandgap ($E_g$), therefore, needs altering the TM-oxygen bonds as well as their interaction energies [15].

First-principles calculations in tetragonal-BFCO yields a narrow $E_g$ ~ 0.78 eV and a large spontaneous polarization, 60.7μC/cm$^2$ [16]. However, such a small $E_g$ is not desirable since it limits the $V_{OC}$ and thus tailoring $E_g$ is critical in further improving conversion efficiency. Nechache et al. accordingly, by tuning the Fe/Cr cation ordering, adjusted the $E_g$ over 1.4-2.4 eV, and demonstrated a conversion efficiency of 8.1% in a multilayer solar cell [17]. Similar bandgap tuning was further reported by Huang et al., who also showed a large spontaneous polarization, 55 μC/cm$^2$, and a high absorption coefficient, $\alpha \sim 2.0\text{-}2.5\times10^5$cm$^{-1}$ [14,18]. Doping lanthanum at the bismuth site, too allowed manipulation of $E_g$ to 1.54 eV within orthorhombic *Pnma* structure, although, optical absorption coefficient is compromised [19]. Nevertheless, an impressive theoretical conversion efficiency, 23.87% was reported in La-doped BFO (BLFO) solar cells, highlighting the material's potential for further optimizations [20]. Thus, tailoring crystal structure through composition and defect engineering is crucial to alter the electronic structure, vis-à-vis optoelectronic properties and therefore PV response in BFCO based materials. In this context, attempt of trying other TMs in the B-site is worth exploring. Mn-based $Bi_2MnCrO_6$ (BMCO) could be a prospective system. Sradhasagar et al. predicted a favorable bandgap in BMCO [9]. Previous experimental work on powder sample prepared through gel-combustion route, demonstrates a monoclinic $P2_1/n$ structure with weak magnetism [21]. However, to the best of our knowledge, no previous attempt to study its detailed optoelectronic characteristics vis-à-vis PV response was undertaken.

In this work, we therefore consider $Bi_2FeCrO_6$ (BFCO) and $Bi_2MnCrO_6$ (BMCO) with monoclinic $P2_1/c$ symmetry (beyond the usual rhombohedral phase) for an in-depth evaluation for prospective PV application. We studied crystal structure, surface morphology, optical properties, and band positions on solution deposited thin films on FTO-coated glass substrates. Consideration of these results enabled solar cell designing and fabrication using films of BFCO / BMCO, buffered between $SnO_2$ and spiro-OMeTAD, electron (ETL) and hole (HTL) transport layers, respectively. Current density-voltage (*J-V*) characteristics of the devices demonstrated conversion efficiencies of 1.07 and 3.56 % for BFCO and BMCO, respectively. *J-V* characteristics were further studied using SCAPS-1D [22]. We predict that a moderate defect density, ~$10^{13}$ cm$^{-3}$ in BMCO could boost the conversion efficiency close to 20 %.

## 2. Methodology

### *2.1 Experimental*

Fluorine-doped tin oxide (FTO) coated glass substrates (TEC7, Sigma Aldrich) were patterned, etched, cleaned and dried (500°C, 15 min.) before thin film deposition. (please refer to the supplementary section for further details). For synthesis of BFCO and BMCO thin films, stoichiometric amounts of the corresponding metal nitrites, *viz.*, $Bi(NO_3)_3.5H_2O$ (Sigma Aldrich ≥ 99.99%), $Cr(NO_3)_3.9H_2O$ (Sigma-Aldrich ≥ 99 %), $Fe(NO_3)_3.9H_2O$ (Sigma-Aldrich ≥ 99.95 %), and $Mn(NO_3)_2{\cdot}xH_2O$ (Sigma-Aldrich ≥ 98 %) were dissolved in a solution of acetic acid, acetyl acetone, and 2-methoxyethanol in appropriate ratios. Thin film was deposited by spin coating at 3000 rpm for 30 s, followed by a sequence of heating at 90 °C, 250 °C, and 450 °C for 5 min. each. The film was subsequently annealed in air for 1 h at an optimized temperature of 650°C to obtain the intended phase. For fabrication of the solar cell, an electron transport layer (ETL) of $SnO_2$ thin film was additionally deposited onto the FTO coated glass substrate by spin coating a 2.5 wt. % solution of $SnCl_4$ in ethanol at 4500 rpm for 30 s, followed by annealing at 500°C for 45 min. for oxide formation. A hole transport layer (HTL) of spiro-OMeTAD, was deposited by spin-coating a solution comprising of 19 μl of bis(trifluoromethylsulfonyl)imide lithium (Li-TFSI) (Sigma–Aldrich) solution (520 mg Li-TSFI in 1 ml acetonitrile) and 21 μl tert-butylpyridine (t-BP) (Sigma–Aldrich) doped in 75 mg of spiro-OMeTAD salt (Sigma–Aldrich) in 1 ml of chlorobenzene at 4500 rpm for 20 s. Finally, silver (Ag) electrode was deposited using shadow mask technique on the Spiro-OMeTAD layer by thermal evaporation.

Crystallographic phase analysis of the films was carried out by powder X-ray diffraction (pXRD) at room-temperature using Cu Kα radiation (λ = 1.5418 Å) in a Bruker D8 Advance diffractometer. The surface morphology was investigated by a Zeiss Merlin Compact field emission scanning electron microscope (FESEM) and atomic force microscope (AFM) (Bruker Nano Wizard Sense, Bio-AFM). Elemental composition was acquired using energy dispersive x-ray spectroscopy (EDS) (Oxford instrument) attached with the FESEM. Optical absorption spectra were acquired using a UV-Visible spectrophotometer (Jasco V-770). X-ray photoelectron spectroscopy (XPS) and ultraviolet photoelectron spectroscopy (UPS) were conducted using Thermo-Scientific NEXSA Surface Analysis with Al K-alpha (1486.6 eV) X-ray source and He plasma at an excitation energy of 21.22 eV. The charge carrier density was determined using Mott-Schottky analysis conducted on a film coated on FTO-coated glass substrate using an electrochemical workstation (Metrohm Autolab PGSTAT204). In an aqueous 0.2 M sodium sulphate ($Na_2SO_4$) solution, a three-electrode system was employed, with Ag/AgCl as the reference electrode and Pt as the counter electrode. The current density-voltage of the device was measured using a source meter (Keithley 2450), integrated with a solar simulator (OREL® LCS-100™), which was calibrated at 1-sun (1000 W $m^{-2}$) AM 1.5 G illumination using a certified mono-Si reference cell (Newport Oriel 2460).

*2.2 Computation: SCAPS simulation*

Frequently used tools for numerical simulation of perovskite solar cells include SCAPS, COMSOL, SETFOS, SILVACO, and ATLAS [23]. (Please refer to Table S1 in the supplementary section for a list of commonly used solar cell simulation tools). The present study employed SCAPS-1D (Solar Cell Capacitance Simulator-One Dimensional, SCAPS 3.3.09), a prevalent tool in photovoltaics research. Although, SCAPS 1D was initially designed for simulating polycrystalline thin-film solar cells, afterwards the capability was enhanced to simulate silicon and perovskite solar cells. The simplicity of modelling a multi-junction perovskite solar cell, consideration of material properties, doping concentrations, energy band structures, and defect densities, are a few key attributes of SCAPS-1D, which performs by solving Poisson's equations, continuity equations, and carrier transport equations for electrons and holes in a one-dimensional scheme [24]. Although SCAPS-1D is heavily used, one must take sufficient precautions to avoid unrealistic results due to inadequate input parameters, and the inherent limitation of not considering optical losses and effective intensity of light not reaching each layers of the device [25]. (more details could be found in the supplementary section) Besides, being a one-dimensional model, SCAPS 1D does not consider effects from 2D and 3D features of the devices such as grain-boundaries, pinholes and surface roughness, for instance. In our study, we have systematically evaluated the accuracy of the input parameters.

Here, we simulate the current-voltage response of the solar cell under solar illumination of AM 1.5 G, considering different defect concentrations, series resistance, and shunt resistance situations. Mimicking the experiments, we chose BFCO and BMCO as the absorber layers, $SnO_2$ and spiro-OMeTAD as the ETL and HTL materials, respectively, whereas FTO-coated glass and Ag served as the front and back contacts. Materials data used in the simulation are given in Tables S2 and S3 in the supplementary section.

## 3. Results and Discussions

*3.1 Structure and morphology*

A comparison of room-temperature powder x-ray diffraction patterns (Fig. 1a) of the films of BFCO and BMCO shows similar peak positions and intensity profiles, indicating that both the compounds demonstrate identical crystallographic phases. Indexing was done using monoclinic $P2_1/c$ (space group number 14) symmetry. A number of double-perovskite systems have been previously indexed using the above monoclinic $P2_1/n$ symmetry [26]. Earlier reports of BFCO and BMCO bulk and thin film samples however demonstrated varied crystal symmetry, *viz.*, monoclinic $P2_1/n$ [21,26], rhombohedral *R3c* [27,28], orthorhombic *Pnma* symmetry [29], and pseudo-cubic [30,31] structures. Fitting the pXRD data using constant scale factor (Le-Bail method) allowed estimation of unit cell parameters in the two compounds, *viz.*, $a$ = 12.8204 Å, $b$ = 5.773 Å, $c$ = 28.8173 Å, $\beta$ = 105.07 (BFCO) and $a$ = 12.7928 Å, $b$ = 5.7310 Å, $c$ = 28.8647 Å, $\beta$ = 105.20 (BMCO).

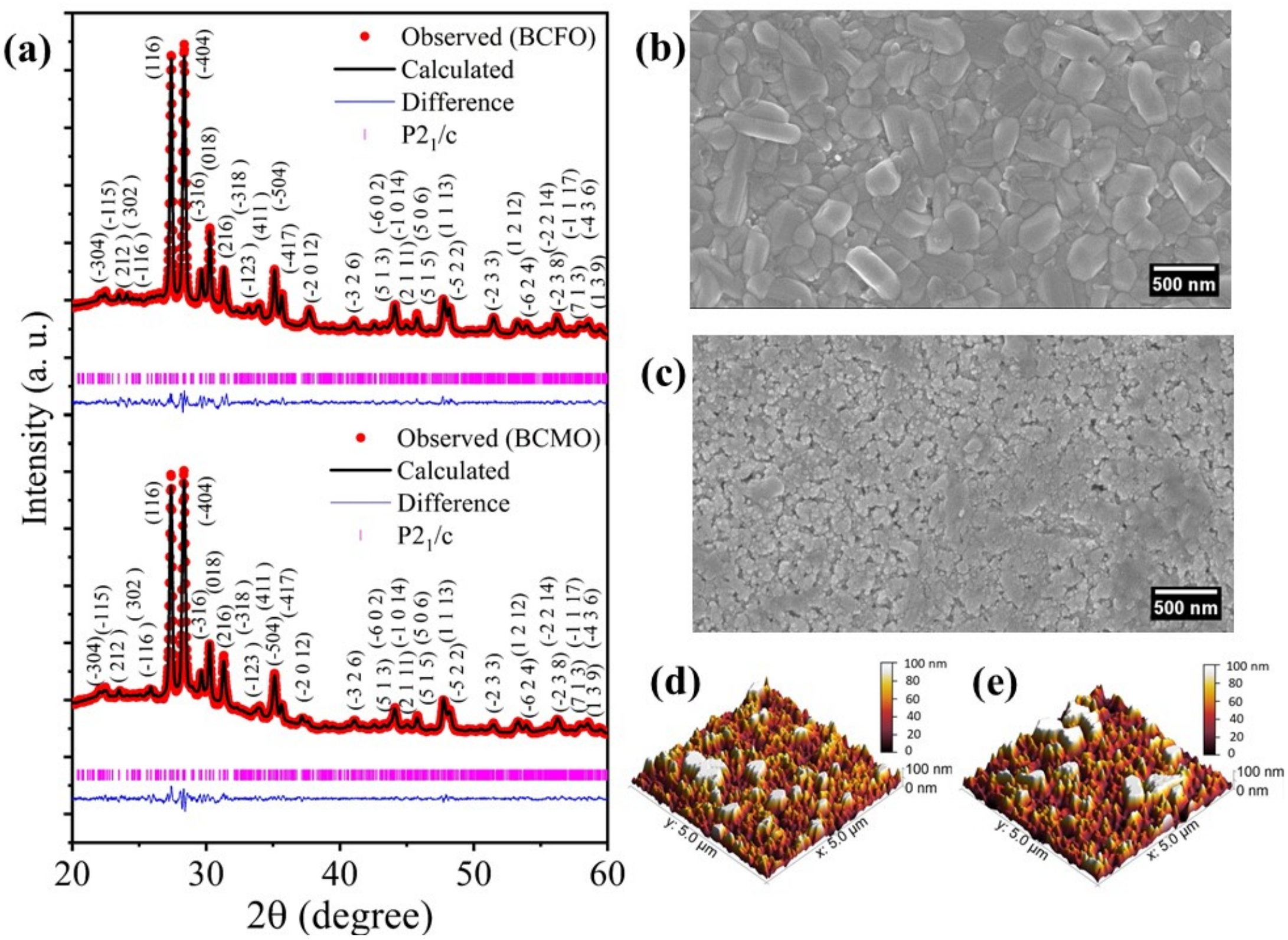

Fig. 1. (a) Room-temperature powder XRD data of BFCO and BMCO films fitted using monoclinic $P2_1/c$ symmetry. Secondary electron FESEM micrographs of (b) BFCO, (c) BMCO thin films deposited on FTO-coated glass substrates. AFM images show surface topography of (d) BFCO, (e) BMCO thin films.

FESEM micrographs (Fig. 1b and 1c) acquired using the in-lens detector exhibit surface morphology of the two films. While BFCO shows a distribution of grain size with smaller grains surrounding a few larger ones, resulting in a fairly compact structure of the films, BMCO shows ultrafine grains coalescing to give rise to clusters. The adjacent clusters share gaps in between, and thus the film is somewhat porous. Corresponding AFM images (Fig. 1d and 1e) demonstrate comparable surface roughness in both BFCO (RMS ~ 20 nm) and BMCO (RMS ~19 nm) films. Favorable orientation vis-à-vis surface energy likely allowed non-uniform growth of fewer grains in the case of BFCO. Cross-sectional SEM images (Supplementary section, Fig. S1) were considered to estimate the film thickness, 346 ± 45 nm (BFCO) and 433 ± 35 nm (BMCO). EDS data confirmed cationic stoichiometry in reasonable extents with those of double-perovskite, *viz.*, 2: 1: 1: 6 (A: B′: B″: O). In the two films, elemental composition demonstrates Bi: Fe: Cr: O ratio ranging over 2.17±0.03: 1.02±0.04: 1.00±0.01: 20.79±0.52 and Bi: Mn: Cr: O ratio, 2.18±0.07: 0.92±0.08: 1.00±0.00: 20.08±0.67 for BFCO and BMCO films, respectively. Higher oxygen content with respect to the stoichiometric composition in the films could be attributed to the FTO coating on the glass substrate. Since the thickness of the absorber layer is small, oxygen pickup from the substrate may have occurred during EDS measurement. A minor contribution from excess Bi-content leading to additional site generation under air ambience, i.e., $\frac{1}{2}O_2 \xrightarrow{Bi} Bi_{Bi} + O_O$ may be further considered. Moreover, oxygen being a lighter element, its x-ray fluorescence yield is inherently low and therefore quantitative estimation of oxygen content in the film using EDS is likely to be less accurate. The slight variation from the cationic stoichiometry may lead to generation of charged point defects responsible for variation in the structure and optoelectronic properties of the films [32].

*3.2 Oxidation states and defect structure*

Room-temperature XPS measurements were performed to investigate the valence states of Bi, Mn/Fe, Cr, and O in BFCO and BMCO films. (Supplementary section, Fig. S2, for surface scan) XPS spectra was calibrated by excitation of the photoelectron corresponding to the C 1s peak centered at 284.6 eV. The perfect fitting of Bi 4f peak with a single curve supports the presence of only the $Bi^{3+}$ oxidation state. (Fig. 2(a),(b)) Cr $2p_{3/2}$ excitation of the spectra corresponding to BFCO, however was fitted with three curves, with peaks centered at 578.3 eV, 576.08 eV and 575.05 eV corresponding to $Cr^{4+}$, $Cr^{3+}$ and $Cr^{2+}$, respectively [21,33]. Corresponding peak positions for $Cr^{4+}$, $Cr^{3+}$ and $Cr^{2+}$ in case of BMCO are 578.45 eV, 575.96 eV, and 574.89 eV, respectively. (Fig. 2(e), (f)) Similarly, two curves were fitted for Fe $2p_{3/2}$ with peaks centered at 711.09 eV and 709.25 eV, corresponding to $Fe^{3+}$ and $Fe^{2+}$ oxidation states of Fe in BFCO [34,35].

(Fig. 2(c)) On the other hand, three curves were fitted for Mn $2p_{3/2}$ at 643.96 eV, 641.16 eV, and 639.90 eV corresponding to $Mn^{4+}$, $Mn^{3+}$, and $Mn^{2+}$ oxidation states in the case of BMCO [36]. (Fig. 2(d)) Multiple oxidation states of the TM-ions (Cr, Fe, and Mn) may lead to formation of charged point defects such as oxygen vacancies as follows, $Cr_{Cr} + O_O \rightarrow \frac{1}{2}O_2 + \ddot{V_O} + C\acute{r}_{Cr} + \acute{e}$; Presence of $\ddot{V_O}$ i.e., loss of oxygen sites alters anion stoichiometry. O 1s state in BFCO (BMCO) could be deconvoluted into three peaks, at 529.36 eV (529.23), 530.6 eV (529.89), and 532.92 eV (532.76), respectively, representing lattice oxygen (Fe-O-Fe/Cr-O-Cr or Mn-O-Mn/Cr-O-Cr ), terminal oxygen (Fe=O/Cr=O or Mn=O/Cr=O), and adsorbed oxygen [34,37].

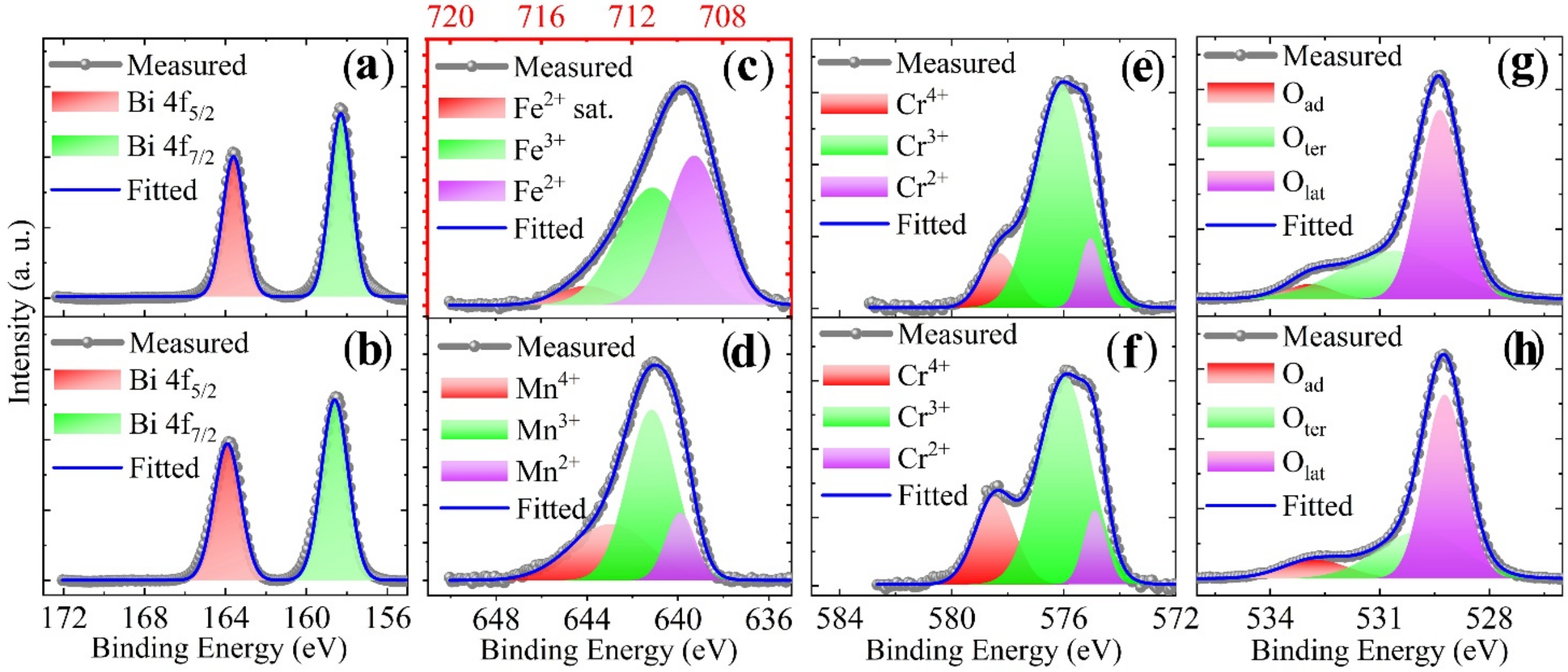

Fig. 2. XPS spectra of BFCO and BMCO film: elemental scans for BFCO: (a) Bi 4f, (c) Fe 2p, (e) Cr 2p, and (g) O 1s; elemental scans for BMCO: (b) Bi 4f, (d) Mn 2p, (f) Cr 2p, and (h) O 1s.

From the EDS data, it was found that both the films have excess bismuth. While BFCO has near stoichiometric transition metal cation composition, BMCO is slightly Mn deficient. Considering the EDS data and XPS analyses, several likely defect scenarios could be proposed; some of which are given in Table 1.

Based on the survey on more than 300 systems, Anderson et al. summarized that, in $A_2B'B''O_6$ double perovskites, long-range ordering at the B site is prevalent when the differences in the formal valence ($\Delta FV_{B'-B''}$) and ionic radii ($\Delta r_{B'-B''}$) of the B-site ions (B'/B") are large [38]. Here, large difference in the ionic radii of the B-site ions ($\Delta r_{B'-B''}$) together with the difference in their charge ($\Delta FV_{B'-B''}$) inhibit site-interchange leading to B-site ordering.

Table 1: Proposed defect reactions in BFCO and BMCO thin films based on EDS and XPS analyses.

| | Likely defect reactions | | |
|---|---|---|---|
| (i) | $null \rightarrow Cr^{\cdot}_{Cr} + Cr'_{Cr}$ | (ii) | $null \rightarrow Mn^{\cdot}_{Mn} + Cr'_{Cr}$ |
| (iii) | $null \rightarrow Cr^{\cdot}_{Cr} + Fe'_{Fe}\,/Mn'_{Mn}$ | (iv) | $Cr_{Cr} + Fe_{Fe}/Mn_{Mn} \rightarrow Cr_{Fe/Mn} + Fe/Mn_{Cr}$ |
| (v) | $\frac{1}{2}O_2 \rightarrow O_O + Cr^{\cdot}_{Cr} + Fe_{Fe}/Mn_{Mn} + \acute{e}$ | | |
| (vi) | $\frac{1}{2}O_2 \rightarrow O_O + Cr^{\cdot}_{Cr} + Mn^{\cdot}_{Mn} + 2\acute{e}$ | (vii) | $\frac{1}{2}O_2 \rightarrow O_O + V'''_{Mn} + Mn^{\cdot}_{Mn} + 2\dot{h}$ |
| (viii) | $Cr_{Cr} + O_O \rightarrow \frac{1}{2}O_2 + \ddot{V}_O + Cr'_{Cr} + \acute{e}$ | (ix) | $Fe_{Fe} + O_O \rightarrow \frac{1}{2}O_2 + \ddot{V}_O + Fe'_{Fe} + \acute{e}$ |
| (x) | $Mn_{Mn} + O_O \rightarrow \frac{1}{2}O_2 + \ddot{V}_O + Mn'_{Mn} + \acute{e}$ | (xi) | $\frac{1}{2}O_2 \rightarrow O_O + V'''_{Mn} + 3\dot{h}$ |
| (xii) | $\ddot{V}_O + \frac{1}{2}O_2 \rightarrow O_O + Cr^{\cdot}_{Cr} + Fe_{Fe} + \dot{h}$ | (xiii) | $\ddot{V}_O + \frac{1}{2}O_2 \rightarrow O_O + Cr^{\cdot}_{Cr} + Mn^{\cdot}_{Mn}$ |

Figure 3 plots $\Delta FV_{B'-B''}$ against $\Delta r_{B'-B''}$ exhibiting regions of disordered, intermediate and ordered double-perovskites. In BFCO and BMCO systems, achieving a completely ordered state with isovalent B-ions is experimentally difficult due to the similar ionic radii and valences (i.e., $Fe^{3+}$, $Cr^{3+}$, and $Mn^{3+}$) of the B-cations (Table 2). It is observed from Table 2 that, one can achieve sufficient $\Delta r_{B'-B''}$ and $\Delta FV_{B'-B''}$ by altering the oxidation states of these ions, viz., $Fe^{2+}$, $Cr^{2+}$, $Mn^{2+}$, $Fe^{4+}$, $Cr^{4+}$, and $Mn^{4+}$. It is to be noted here that octahedral coordination of $Cr^{2+}/Mn^{2+}/Fe^{2+}$ -ions with $O^{2-}$-ions generally involves high-spin configuration of the TM ions. Acquiring different oxidation states by the B' and B"-ions than FV= +3, would involve different defect reaction scenarios involving creation and annihilation of oxygen vacancies, $\ddot{V}_O$, as shown in (viii) to (xiii) in Table 1. Defect reactions (viii)-(x) in Table 1, involving creation of $\ddot{V}_O$ requires an oxygen deficient ambient while those in (xi)-(xiii) in Table 1, demand oxygen-rich conditions. Hence, an optimal oxygen partial pressure is essential to balance the formation of an even concentration of both types of charged defects. Oxygen vacancies may act as mid-gap states promoting SRH recombination vis-à-vis efficiency loss. Therefore, maintaining an optimum oxygen partial pressure is critical.

Table 2: Oxidation states and ionic radii of the constituent B-site transition metal ions in BFCO and BMCO.

| Cr: [Ar]$3d^5 4s^1$ | | Mn: [Ar]$3d^5 4s^2$ | | Fe: [Ar]$3d^6 4s^2$ | |
|---|---|---|---|---|---|
| Ion$^{oxid.state}$ | Ionic rad. (Å) | Ion$^{oxid.state}$ | Ionic rad. (Å) | Ion$^{oxid.state}$ | Ionic rad. (Å) |
| $Cr^{2+}$:[Ar]$3d^4$ | 0.73 (low spin) | $Mn^{2+}$:[Ar]$d^5$ | 0.67 (low spin) | $Fe^{2+}$:[Ar]$3d^6$ | 0.61 (low spin) |
| | 0.80 (high spin) | | 0.83 (high spin) | | 0.78 (high spin) |
| $Cr^{3+}$:[Ar]$3d^3$ | 0.615 | $Mn^{3+}$:[Ar]$3d^4$ | 0.58 (low spin) | $Fe^{3+}$:[Ar]$3d^5$ | 0.55 (low spin) |
| | | | 0.645 (high spin) | | 0.645 (high spin) |
| $Cr^{4+}$:[Ar]$3d^2$ | 0.55 | $Mn^{4+}$:[Ar]$3d^3$ | 0.53 | $Fe^{4+}$:[Ar]$3d^4$ | 0.585 |

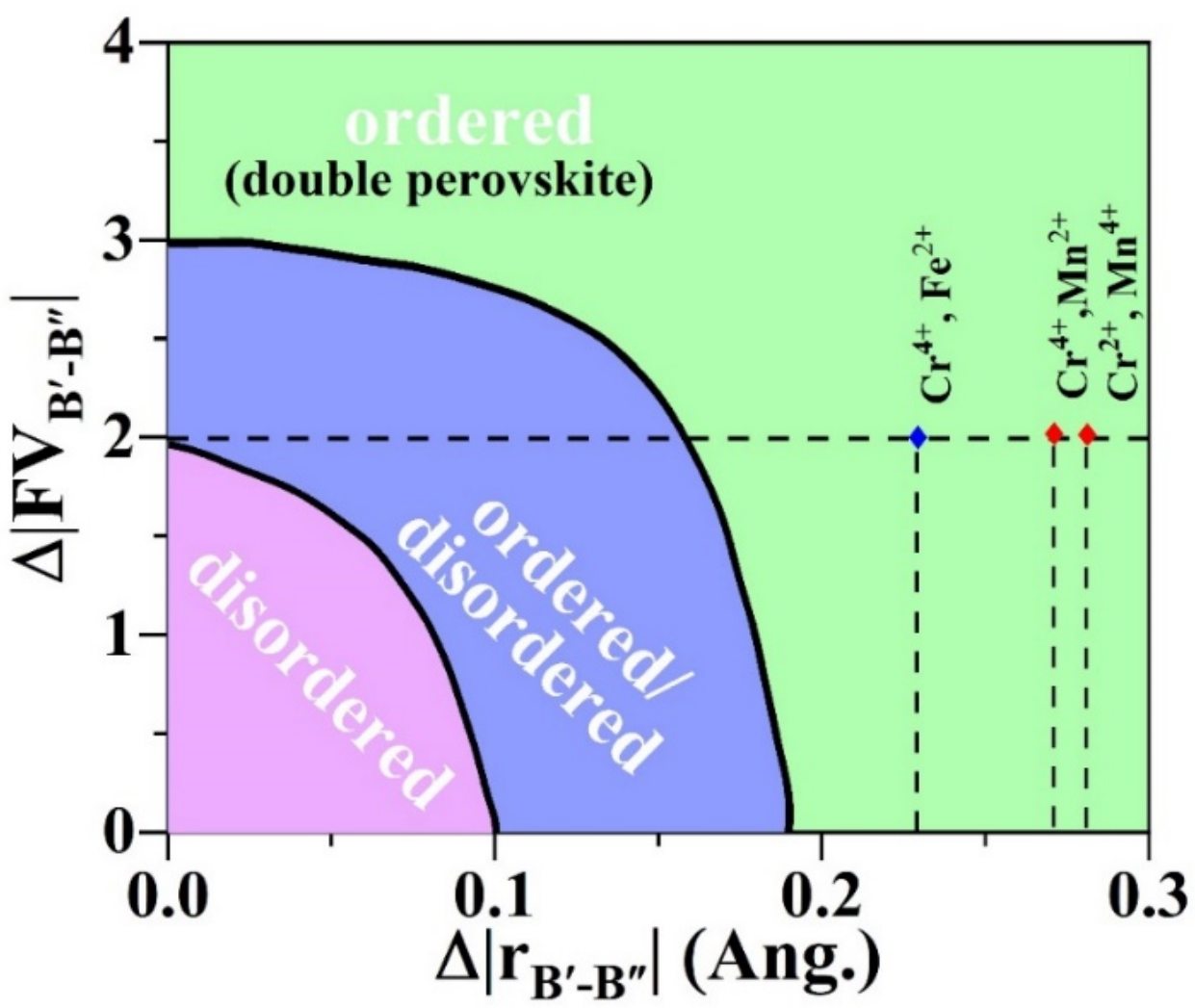

Fig. 3: $\Delta FV_{B'-B''}$ vs $\Delta r_{B'-B''}$ plot in relation with BFCO and BMCO. (adapted from the work of Anderson et al. [38])

Nechache et al., in their work on BFCO reported that kinetics of film growth was also important in achieving long-range ordering of perovskite structure [17]. Ordered domain size (D) and the long-range order (LRO) parameter, R (ratio of intensities of (½ ½ ½) superstructure reflection and (111) fundamental perovskite peak) are dependent on growth rate and substrate temperature [17]. Slow growth rate and higher temperature enable long-range ordering and fewer anti-site defects. Such controlled cation ordering (obtained through optimization of the growth conditions) can modify the bond length and energy, and thereby allow tuning the material's bandgap. In their work Nechache et al. showed that formation of $Fe^{2+}$ and $Cr^{4+}$ oxidation states follow the criteria of $\Delta FV_{B'-B''}$ and $\Delta r_{B'-B''}$, enabling B-site ordering that allows tailoring the bandgap of the system [17].

In the present work, BFCO and BMCO are associated with the presence different oxidation states. XPS analysis show that BFCO contains $Fe^{2+}$ (50.2%), $Fe^{3+}$ (49.8%), $Cr^{2+}$ (9.92%), $Cr^{3+}$ (77.53%), and $Cr^{4+}$ (12.55%). Similarly, BMCO has $Mn^{2+}$ (13.67%), $Mn^{3+}$ (56.06%), $Mn^{4+}$ (3.27%), $Cr^{2+}$ (23.8%), $Cr^{3+}$ (56.18%), and $Cr^{4+}$ (23.8%). In our work, BFCO does not possess $Fe^{4+}$ states. Therefore, $Fe^{2+}$ and $Cr^{4+}$ states in BFCO can enhance cation ordering. In BFCO, a low fraction of $Cr^{4+}$ and the absence of $Fe^{4+}$ are not favourable for an ordered structure; In case of BMCO, however, presence of $Mn^{4+}$, $Mn^{2+}$, $Cr^{2+}$, and $Cr^{4+}$ ensure a superior B-site ordering possibility. The required oxidation states can be achieved by treating films in an optimum oxygen partial pressure environment with slow heating rate that may promote cation diffusion, increase domain size, and reduce antisite disorder, thereby improving film quality.

Another approach is to reduce defects through passivation engineering akin to the more mature field of lead-halide perovskite solar cells. Here, mono-/di-/tri-valent metal ions doping could be used to passivate negatively charged defects of different charges as well as to reduce anti-site defects of the TM-ions [39]. For instance, $Na^+$ and $K^+$-doping are typically aimed at passivating negatively charged metal vacancies on the surface and at the grain boundaries of lead-halide perovskites by forming ionic-bonds with electronegative halogen ions [40,41]. Controlled $Na^+$-doping increases grain-size and thereby reducing pinhole concentration [42,43]. $K^+$-ions are known to eliminate hysteresis effect by minimizing anti-Frenkel pair formation [40]. Optimum concentration of $K^+$-ions are also shown to passivate the uncompensated halogen ions at the grain boundaries [44]. Doping with small amount of metal-ions with larger charges such as $Al^{3+}$, $In^{3+}$, $Sb^{3+}$ are claimed to reduce micro-strain in the films and thus, reduce defect density. Furthermore, passivation with Lewis acids and bases have also been demonstrated in lead halide perovskite systems. Passivating BFCO and BMCO using similar approaches is an open area of research. Furthermore, to improve the interface compatibility, an ultra-thin, buffer layer ~1-5 nm of systems such as $Al_2O_3$ or $SiO_2$ between ETL and Perovskite may be added that can modify the grain boundary defects [45], boosting the overall performance of the device.

### *3.3 Optical properties*

Fig. 4a plots the optical absorption spectra of BFCO and BMCO films on FTO-coated glass substrates. The absorption spectra of the films demonstrate an exponential behavior with incident photon energy having a maximum absorption coefficient of the order of ~ $10^4$-$10^5$ $cm^{-1}$ at around 2.5 eV, indicating that the main portion of the solar spectrum is absorbed by the materials under study, a crucial attribute for PV application [46]. BMCO demonstrates a higher α value at all values of incident photon energy. Furthermore, absorption spectra depict free-exciton peaks at ~3.09 eV for BMCO and at ~3.11 eV for BFCO, respectively.

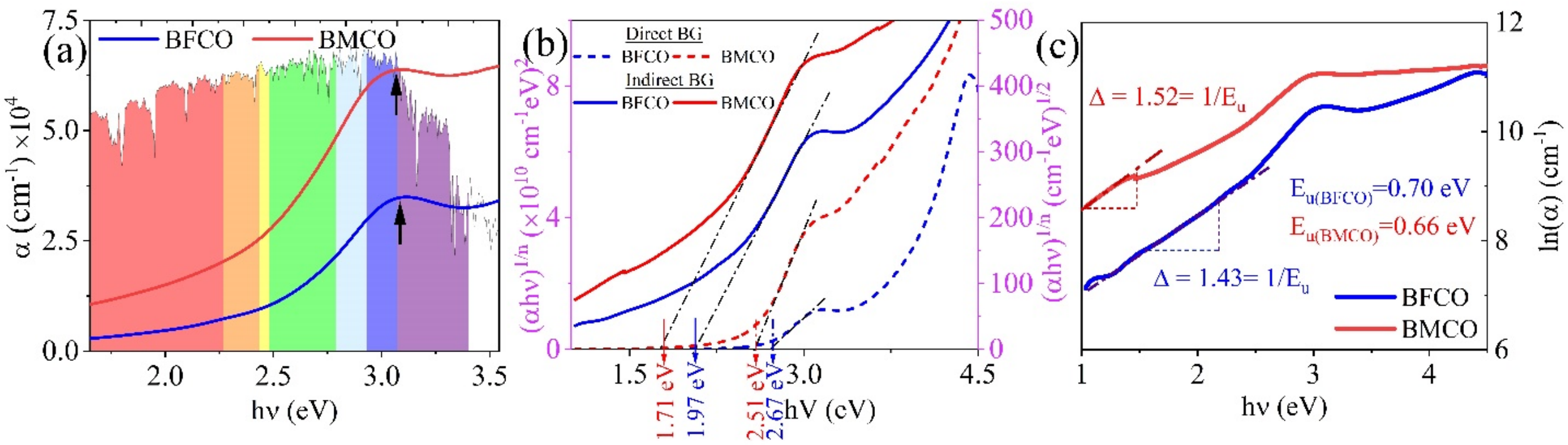

Fig. 4. (a) Optical absorption spectra of BFCO and BMCO films, (b) Tauc plot for direct band gap and indirect band gap for BFCO and BMCO (c) Urbach energy calculation.

The bandgap of the materials was estimated using the Tauc equation [47].

$$\alpha h\nu = ( h\nu - E_g)^n \quad (1)$$

Where $E_g$ is the bandgap, and the exponent, $n$ is a dimensionless parameter equal to ½ and 2 for the direct and indirect transitions, respectively. The bandgap is determined by the $x$-axis intercept of the linear portion of the $(\alpha h\nu)^n$ versus $h\nu$ plot. Fig. 4b suggests that both the materials demonstrate indirect bandgap characteristics with the estimated indirect bandgap, 1.97 and 1.71 eV, for BFCO and BMCO, respectively. Earlier reports on the characteristics and magnitude of bandgap in BFCO has been diverse. Huang *et al.* reported a direct band gap of ~ 1.5 eV [34]. Nechache et al. [17], demonstrated band gap tuning from 1.5-2.2 eV by varying the Fe/Cr ratio, while Quattropani et al. [48], reported a tunable band bap in the range of 1.9-2.6 eV clearly demonstrating the role of TM-3d state in forming the VB and CB. Octahedral coordination of oxygen with Fe and Cr form [$FeO_6$] and [$CrO_6$] polyhedra in BFCO; the distorted octahedra cause the splitting of 3d (in Fe and Cr) orbital states into low energy state $t_{2g}$ (*i.e.*, $d_{xy}$, $d_{yz}$ and $d_{zx}$) and high energy $e_g$ states ($d_{x^2-y^2}$ and $d_{z^2}$). The valance band edge is formed by overlapping Cr 3d, Bi 6s and O 2p, whereas the conduction band edge is made up with spin up O 2p, Fe 3d and Cr 3d states, and spin down Bi 6p and O 2p [31,49]. It must be noted that the relative position and composition of VB and CB depend on the crystallographic symmetry. Bhardwaj et al. [26], reported that BMCO with Bi: Mn: Cr: O ratio of 1:0.9:0.1:3 has a bandgap of 1.25 eV, which is lower as compared to other data. Sradhasagar et al. [9], predicted a bandgap of 1.57 eV in the stoichiometric composition within *R3* symmetry. Senthilkumar et al. [21], however reported an indirect bandgap of 2.35 eV in BMCO powder with monoclinic $P2_1/n$ symmetry. Thus, the bandgap of BMCO also depends on the Mn/Cr ratio as in the case of BFCO [17]. Antisite defects are formed due to the disordering of Fe/Cr in the BFCO, and Mn/Cr in BMCO that play crucial role in tuning the band gap [50].

Apart from the normal band-to-band excitation from the valence band to the conduction band, defect mediated transition is also feasible [51]. A transition from a localized tail state, present just above the valence band to another localized tail state, below the conduction band is possible [52]. The extended tail into the bandgap, is termed Urbach tail and the corresponding energy, is referred to as the Urbach energy ($E_U$), where is $E_U$ can be calculated from [53]:

$$\alpha(h\nu) = \alpha_0 exp\left(\frac{h\nu - Eg}{E_U}\right) \quad (1)$$

here $\alpha$ is the absorption coefficient, $h$ is the Planck constant, $\nu$ is the frequency of light. As shown in Fig 4c, $E_U$ may be obtained by taking the reciprocal of the slope of $ln\ \alpha$ v*s.* $h\nu$ plot. In literature, Urbach tails have been linked to phonons, impurities, excitons/ electron-phonon coupling, and structural disorders in the materials [54–57]. $E_U$ affects electrical properties such as carrier mobility, lifetime, absorbance, saturation current, and radiative recombination pathways, including band-to-band, band-to-tail, and tail-to-band. Radiative transitions between localized states, such as tail-to-tail, band-to-tail, or tail-to-band, lead to Stokes shifts. A large value of $E_U$ indicates a deep defect state, which limits the actual $V_{OC}$ of the device and increases the $V_{oc,deficit}$ ($V_{oc,deficit}$ = $E_g$ - $V_{OC}$). The deep defect state facilitates Shockley–Read–Hall (SRH) recombination, and thus the $V_{OC}$ decreases with increasing depth of the defect level, $E_t$ in the forbidden gap. Whereas $J_{SC}$ remains nearly constant for shallow defect levels ($E_t$<0.2 eV), it falls steeply for deep ($E_t$>0.3 eV) defect levels [52]. Previous study on vapor deposited films of $MAPbI_3$ reported an $E_t$ of 0.05 eV translating to shallow defect state with few structural and thermal disorders [53]. For BFCO and BMCO, the estimated $E_U$ are 0.7 eV and 0.66 eV, respectively. Thus, the depth of defect level, $E_t$, for BFCO and BMCO are 0.35 eV and 0.33 eV, respectively suggesting deep defect level ($E_t$ > 0.3eV) in both the films [52].

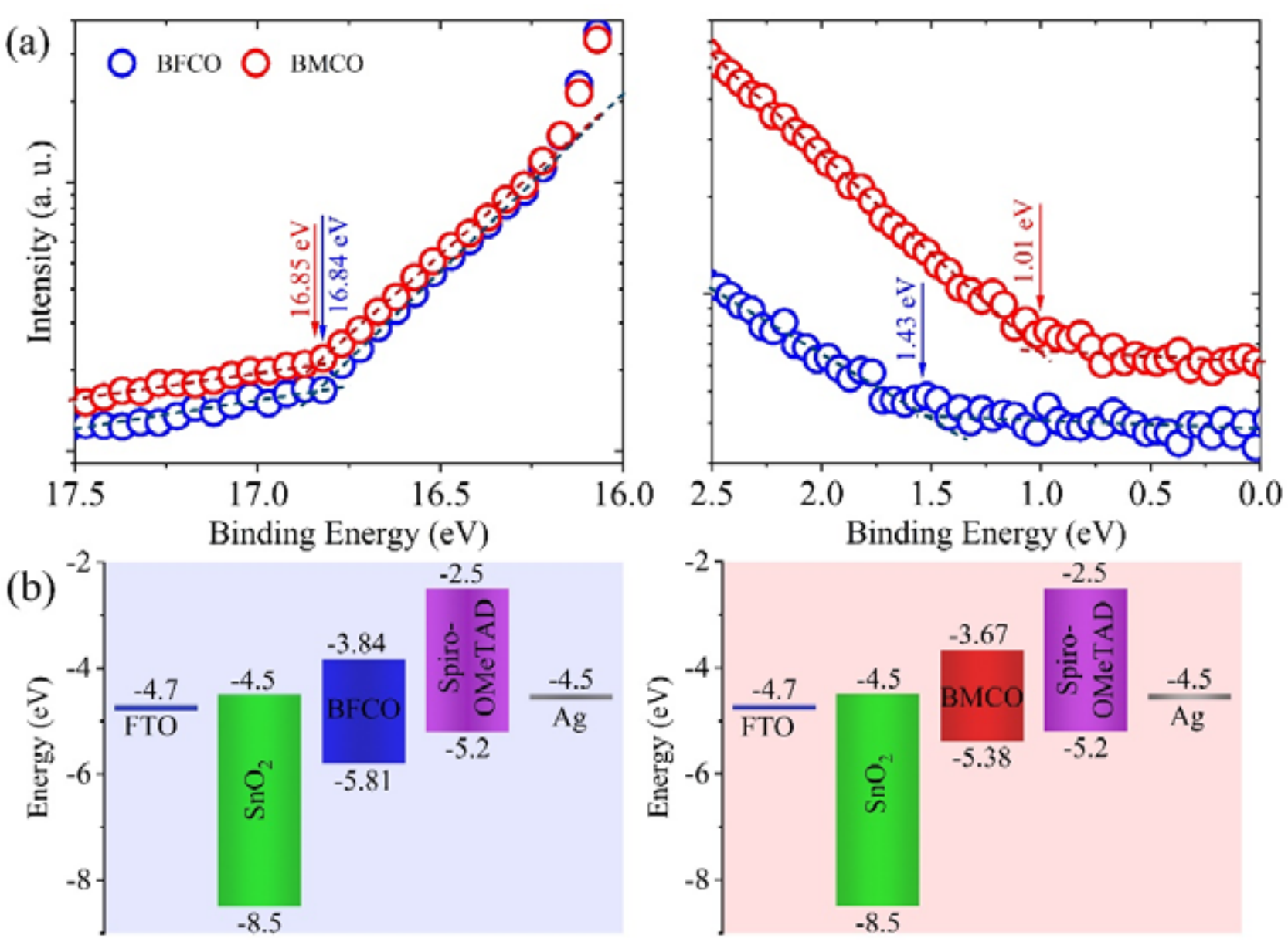

Fig. 5. Ultraviolet photoelectron spectroscopy data of BFCO and BMCO thin films (a) secondary energy cut-off energy; onset electron energy of BFCO and BMCO on the specified range. (b) Schematics showing band energy alignment of BFCO, BMCO, charge transport layers, and electrodes.

In a perovskite solar cell, excitons are generated in the absorber layer as a result of photon energy exceeding the band gap of the material. These excitons are separated into independent charge carriers due to the difference in work functions between the electrode layers and the positions of the conduction and valence band edges of the absorber layer [58]. Further charge carriers are extracted to external circuits. Thus, for photovoltaic device fabrication, the knowledge of the position of band edges is critical, accordingly the band edges of the charge transfer layers and electrodes are aligned such that the generated exciton, on exposure to light can be separated, and electrons/holes are efficiently drained to the external circuits. Fig. 5a shows the ultraviolet photoelectron spectroscopy (UPS) spectra of BFCO and BMCO, from which the Fermi level ($E_F$) can be determined as, $E_F = 21.22 - E_{SECO}$, where $E_{SECO}$ is the secondary electron cut-off energy. (Fig. 5a) Thus, for BFCO and BMCO, the Fermi energies are 4.38 eV and 4.37 eV, respectively, where there is a rise in the Fermi level for BMCO as compared to BFCO. The valence band maximum ($E_{VBM}$) of the absorber layer can be calculated using, $E_{VBM} = 21.22 – (E_{SECO} – E_{onset})$, where $E_{onset}$ is the onset electron energy [59]. The $E_{SECO}$ and $E_{onset}$ are determined from the intersection of the lines through the linear portions of the initial incremental and the final decremental curves with the adjacent linear portions of the curves, respectively, as shown in Fig.

5a, where $E_{onset}$ is the difference between the Fermi level and the valence band maxima. The conduction band minimum and valence band maximum for BFCO (and BMCO) are 5.81 eV (5.38 eV) and 3.84 eV (3.67 eV), respectively. The conduction band minimum was obtained by adding bandgap energy to the valence band maximum, obtained from UV-Vis spectroscopy. Accordingly, ETL and HTL materials were selected, $SnO_2$ and Spiro-OMeTAD [60]. Fig. 5(b) schematically shows relative positions of the band edges of BFCO and BMCO relative to those of ETL ($SnO_2$), HTL (Spiro-OMeTAD) and the electrodes.

*3.4 Mott-Schottky measurement*

The effective charge carrier density ($N_D$), was further estimated using Mott-Schottky measurement assuming ideal semiconductors,

$$N_D = \frac{2}{e\varepsilon\varepsilon_0}\left[\frac{d\left(1/C^2\right)}{dV}\right]^{-1} \quad (2)$$

Where $e$ is the charge of the electron ($1.6 \times 10^{-19}$ C), $C$ is the space charge capacitance, $V$ is the bias voltage, $\varepsilon$ and $\varepsilon_0$ are relative permittivity of the material, and vacuum permittivity ($8.85 \times 10^{-14}$ F/cm), respectively. $\frac{d\left(1/C^2\right)}{dV}$ is the slope of the Mott-Schottky curve, n-type semiconductor has a positive slope, and p-type semiconductor shows a negative slope characteristics [61].

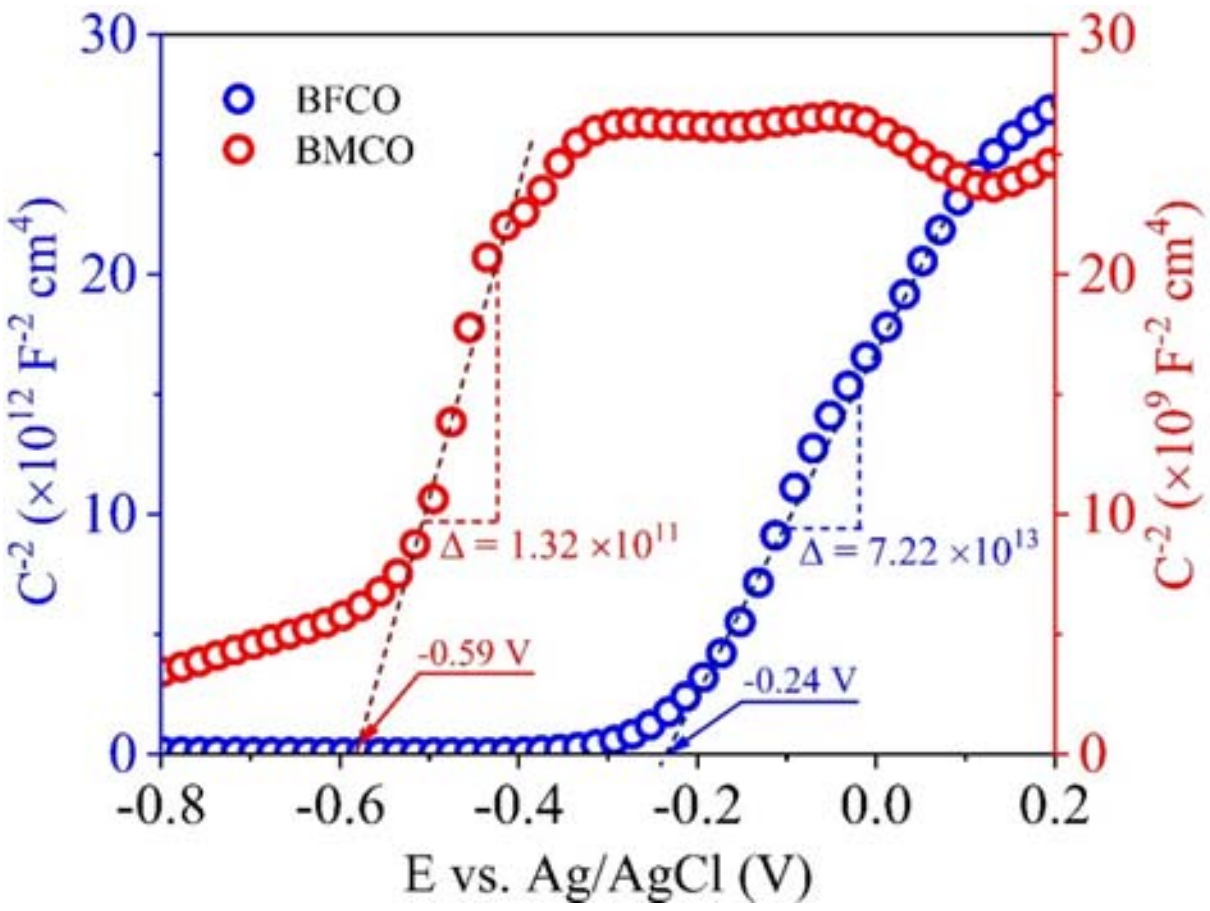

Fig. 6. Mott-Schottky analysis of BFCO and BMCO films on FTO-coated glass substrates.

The slope of $1/C^2$ vs. V plot (Fig. 6) for BFCO ($7.22 \times 10^{13}$) is greater than BMCO ($1.32 \times 10^{11}$) and both the materials behave as n-type material. Corresponding charge carrier densities are $1.59 \times 10^{17}$ cm$^{-3}$ and $1.07 \times 10^{20}$ cm$^{-3}$, respectively. It was noted that the charge carrier density of BMCO is greater than that of BFCO, which render BMCO to perform better as an absorbing material. The above attributes were considered for simulating the current-voltage (*J-V*) characteristics of BFCO and BMCO based perovskite solar cells using SCAPS-1D.

*3.5 Photovoltaic measurements*

Fig. 7 shows the *J-V* characteristics of the solar cells fabricated with BFCO and BMCO as absorber layers. The device parameters of the two devices, *viz.*, FTO/$SnO_2$/BFCO/Spiro-OMeTAD/Ag (Device-1) and FTO/$SnO_2$/BMCO/Spiro-OMeTAD/Ag (Device-2) are given in Table 3. The power conversion efficiency of Device-2 is higher than that of Device-1, and Device-2 exhibits a higher fill factor, which may be attributed to improved band alignment, larger absorption coefficient, and a higher charge carrier density in BMCO. However, it is observed that in both cases, the characteristics of the *J-V* curve deviates from the usual exponential behavior under illumination. Features like, red kink and crossover were observed [62,63]. The open circuit voltage ($V_{OC}$) is limited to 0.69 V for BFCO and 0.67 V for BMCO, respectively. This type of behavior could be attributed to defects in the absorbing layer, as shown earlier in sections 3.2 and 3.3, resulting in variation in the *J-V* curve behavior and thereby limiting the device efficiency [63]. A similar type of *J-V* behavior was reported by Chung et al. for a solution-processed high-performance CuIn(S, Se)$_2$ solar cell; the red kink and crossover appear due to non-Ohmic back-contact, deep-level defects in the absorber layer, and at the interface of the charge transport layers [64]. In addition, shunt resistance and series resistance also play crucial roles in the variation of *J-V* behavior [58].

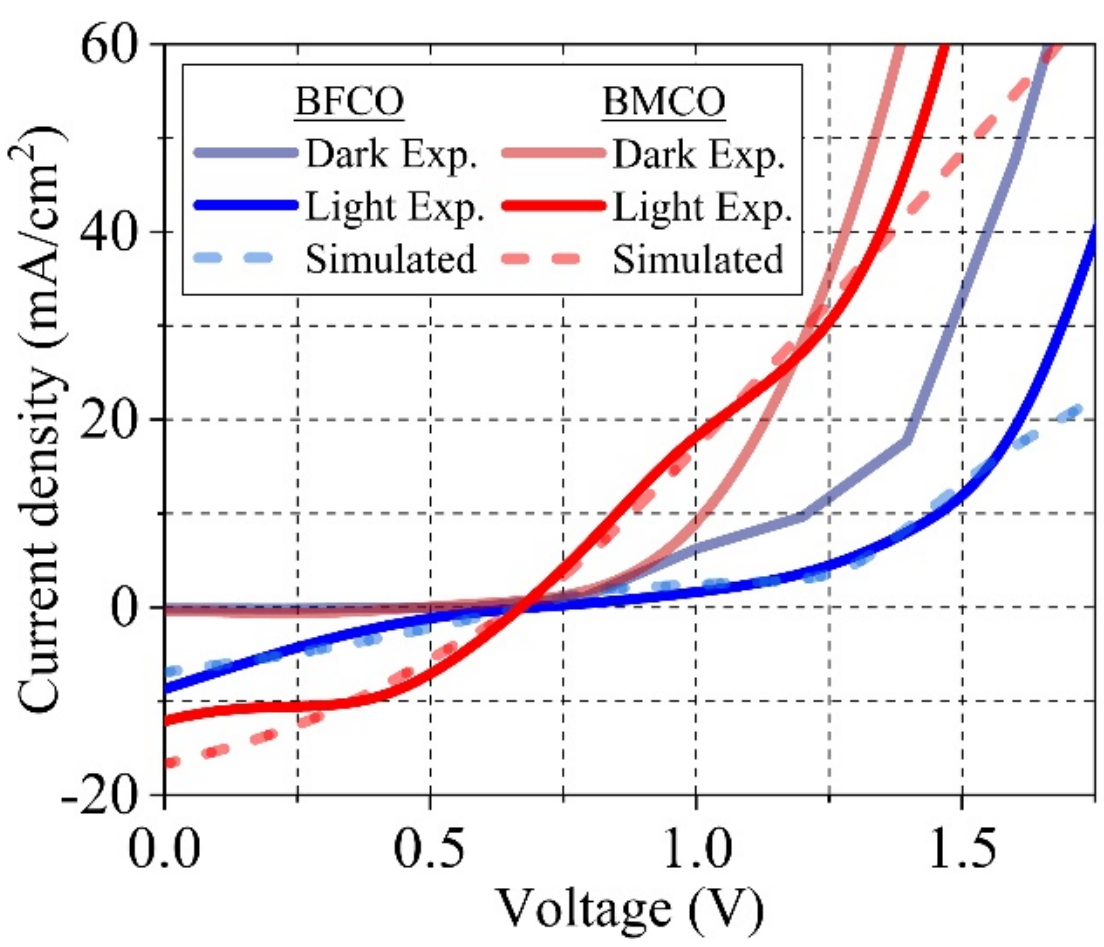

Fig. 7. Experimental J-V curve of Devices 1 and 2 at dark and light. Simulated J-V curve for Devices 1 and 2 under illumination.

To rationalize the measured *J-V* characteristics, we further simulated the *J-V* plot one dimensional drift diffusion software SCAPS 1D [65], which is based on solving Poisson's and continuity equation.

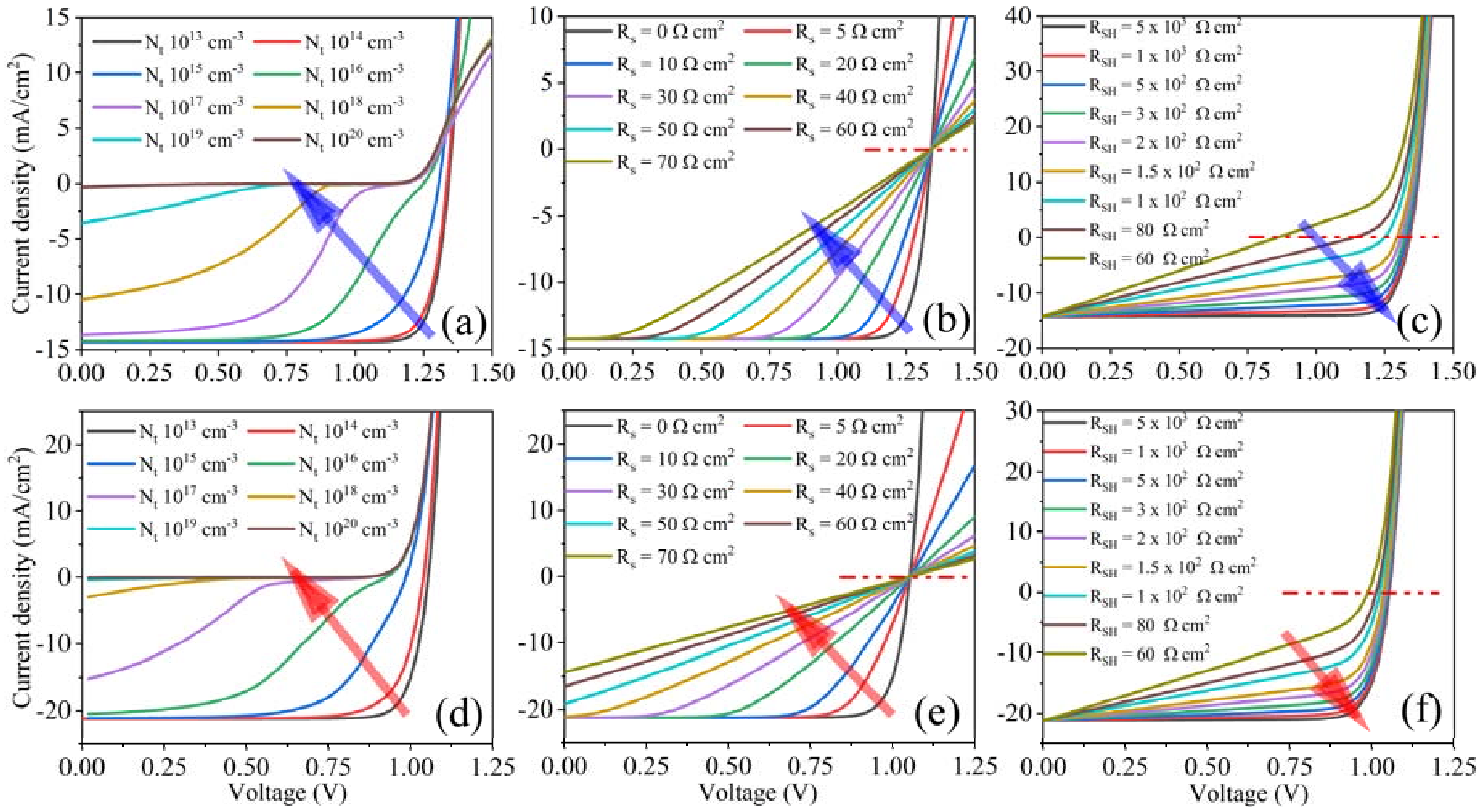

Fig. 8. *J-V* curves of the BFCO-based device modelled by SCAPS-1D software by varying the (a) total defect density ($N_t$), (b) series resistance ($R_S$), and (c) shunt resistance ($R_{SH}$), respectively; (d), (e) and (f) corresponding *J-V* plots for the BMCO-based device.

Fig. 8 shows the *J-V* characteristics of Device 1 (FTO/$SnO_2$/BFCO/Spiro-OMeTAD/Ag) and Device 2 (FTO/$SnO_2$/BMCO/Spiro-OMeTAD/Ag) simulated considering AM 1.5, illumination with a solar irradiation of 100 mW/cm$^2$ ($P_{in}$). The input parameters related to each layer and interlayer defect are given in Tables S2 and S3 in the supplementary section. Here, we study the behavior of *J-V* by varying only one parameter at a time. The total defect density of the absorbing layer is varied from $10^{13}$ cm$^{-3}$ to $10^{20}$ cm$^{-3}$, causing an S-shaped *J-V* curve for both the devices, as evident from Fig. 8a and 8d. The defects and misalignment of band energy at the interface cause an increase in charge accumulation, and the gap state recombination current generated affects the output current, altering the shape of the *J-V* curve from the regular exponential shape that photovoltaic devices typically exhibit [66]. However, the crossover of the *J-V* curve occurs due to hysteresis, which is dependent on surface recombination and the carrier diffusion length [62]. Series resistance in cell arises from contact resistance between each layer and the metal contact, which directly affects the $J_{SC}$. From fig. 8b and 8e, an increasing $R_S$ causes a decrease in $J_{SC}$. Shunt resistance arises due to defects present in the system, causing a decrease in the $V_{OC}$. A similar study on Eu (0%, 4%, 8% and 12%) doped $BiFeO_3$-based perovskite systems, is comparable to our results [58].

The simulated open circuit voltage ($V_{OC}$), short circuit current density ($J_{SC}$), maximum current density ($J_{MAX}$), and maximum voltage ($V_{MAX}$), along with fill factor ($FF$) and power conversion efficiency ($\eta = (V_{OC} \times J_{SC} \times FF)/P_{in}$) for both BFCO and BMCO-based devices, are given in Table S5 (supplementary section). From Fig. 8, it is evident that $N_t$, $R_S$, and $R_{SH}$ affect the behavior of the *J-V* curve. In the subsequent study, these parameters are optimized such that the experimental *J-V* curves are reproduced in the simulation to a reasonable extent. Fig. 7 also shows a comparison of the experimental and simulated *J-V* characteristics of devices 1 and 2 by tuning the input parameters. The optimum values of $N_t$, $R_S$, and $R_{SH}$ for BFCO are $2 \times 10^{18}$ $cm^3$, 10 $\Omega.cm^2$ and 400 $\Omega.cm^2$, respectively and for BMCO, are $2 \times 10^{16}$ $cm^{-3}$, 15 $\Omega.cm^2$ and 140 $\Omega.cm^2$ respectively that reproduce experimental *J-V* characteristics.

Table 3: Performance of BFCO and BMCO based devices (experiments and simulation data)

| Parameters | Exp. BFCO | Sim. BFCO | Exp. BMCO | Sim. BMCO |
|---|---|---|---|---|
| $V_{oc}$ (V) | 0.69 | 0.67 | 0.67 | 0.67 |
| $J_{sc}$ ($mA/cm^2$) | 8.78 | 6.91 | 12.15 | 17.01 |
| FF (%) | 18 | 29.25 | 44 | 34.16 |
| η (%) | 1.07 | 1.35 | 3.56 | 3.92 |
| $V_{max}$ (V) | 0.25 | 0.36 | 0.5 | 0.379 |
| $J_{max}$ ($mA/cm^2$) | 4.27 | 3.77 | 7.11 | 10.33 |

Extrapolation of the predicted device parameters with the consideration of better thin film process optimization, leading to $N_t$, $R_S$, and $R_{SH}$ values $10^{13}$ $cm^{-3}$, 0 $\Omega.$ $cm^2$, and $5\times10^3$ $\Omega.$ $cm^2$, results in ~ 19 % conversion efficiency in the case of BMCO and ~ 17% for BFCO, respectively. Such promising results demonstrate the potential of the materials and justify the requirement of further optimization of the film deposition conditions to produce an optimum defect concentration vis-à-vis improved device performance.

**Conclusions**

Thin films of bismuth-based double-perovskites BFCO (~350 nm) and BMCO (~430 nm) were solution-deposited on FTO-coated glass substrates. Room-temperature powder x-ray diffraction demonstrates monoclinic $P2_1/c$ symmetry in both the materials. Energy dispersive spectroscopy showed excess bismuth, although overall cationic stoichiometry represents those of double-perovskites. From the XPS and EDS data possible point defects were proposed. Optical absorption measurement showed that BFCO exhibits an absorption coefficient of $3.58 \times 10^{-4}$ $cm^{-1}$, with an indirect bandgap of 1.97 eV, whereas BMCO shows a higher absorption coefficient of $6.39 \times 10^{-4}$ $cm^{-1}$ and a narrower band gap of 1.71 eV. The calculated Urbach energy confirmed the presence of deep level defect-related localized states in both materials. UPS revealed the positions of VBM and CBM in the two systems, thus helped in choosing the carrier transport layer materials for solar cell fabrication. Carrier densities, $1.59 \times 10^{17}$ $cm^{-3}$ and $1.07 \times 10^{20}$ $cm^{-3}$ in BFCO and BMCO were estimated using Mott-Schottky measurements. *J-V* characteristics of the solar cells yielded power conversion efficiencies of ~1.07% and ~3.56%, respectively, although presence of red kink and cross-over were noted. Numerical simulation using SCAPS-1D predicted the critical role of defect density in bringing down the device performance, and suggests the requirement of process optimization. It was shown that optimum defect concentration can boost the device performance close to 20%.

**CRediT authorship contribution statement**

**N P Vikas:** Methodology, Data curation, Data analysis, writing – first draft, editing, Software. **Ranjit K Pradhan:** Data curation and analysis; **Somdutta Mukherjee:** Formal analysis, writing and review, **Uday P Singh:** writing and review, **Biplab K Patra:** writing and review, **Ravi P Srivastava:** Methodology, Formal analysis, writing and review, **Amritendu Roy:** Writing – review & editing, Supervision, Software, Resources, Funding acquisition, Formal analysis, Conceptualization.

**Declaration of Competing Interest**

The authors declare that they have no known competing financial interests or personal relationships that could have appeared to influence the work reported in this paper.

**Acknowledgement**

The work was partially funded by Science and Engineering Research Board, Department of Science and Technology, Govt. of India through project no. CRG/2019/003828.

## References

[1] W.-J. Yin, B. Weng, J. Ge, Q. Sun, Z. Li, Y. Yan, Oxide perovskites, double perovskites and derivatives for electrocatalysis, photocatalysis, and photovoltaics, Energy Environ. Sci. 12 (2019) 442–462. https://doi.org/10.1039/C8EE01574K.

[2] S. Walia, S. Balendhran, H. Nili, S. Zhuiykov, G. Rosengarten, Q.H. Wang, M. Bhaskaran, S. Sriram, M.S. Strano, K. Kalantar-zadeh, Transition metal oxides – Thermoelectric properties, Progress in Materials Science 58 (2013) 1443–1489. https://doi.org/10.1016/j.pmatsci.2013.06.003.

[3] M.G. Basavarajappa, S. Chakraborty, Rationalization of Double Perovskite Oxides as Energy Materials: A Theoretical Insight from Electronic and Optical Properties, ACS Materials Au 2 (2022) 655–664. https://doi.org/10.1021/acsmaterialsau.2c00031.

[4] S.R. Basu, L.W. Martin, Y.H. Chu, M. Gajek, R. Ramesh, R.C. Rai, X. Xu, J.L. Musfeldt, Photoconductivity in BiFeO3 thin films, Applied Physics Letters 92 (2008) 091905. https://doi.org/10.1063/1.2887908.

[5] S. Rahmany, L. Etgar, Semitransparent Perovskite Solar Cells, ACS Energy Letters 5 (2020) 1519–1531. https://doi.org/10.1021/acsenergylett.0c00417.

[6] A.K. Astakala, S.-Y. Lee, J. Gautam, K.B. Thapa, I. In, S.J. Lee, S.-J. Park, Engineering inorganic perovskite solar cells: overcoming efficiency and stability barriers for next-generation photovoltaics, Advanced Powder Materials 5 (2026) 100354. https://doi.org/doi.org/10.1016/j.apmate.2025.100354.

[7] M. Caputo, N. Cefarin, A. Radivo, N. Demitri, L. Gigli, J.R. Plaisier, M. Panighel, G. Di Santo, S. Moretti, A. Giglia, M. Polentarutti, F. De Angelis, E. Mosconi, P. Umari, M. Tormen, A. Goldoni, Electronic structure of MAPbI3 and MAPbCl3: importance of band alignment, Scientific Reports 9 (2019) 15159. https://doi.org/10.1038/s41598-019-50108-0.

[8] R.E. Brandt, V. Stevanović, D.S. Ginley, T. Buonassisi, Identifying defect-tolerant semiconductors with high minority-carrier lifetimes: Beyond hybrid lead halide perovskites, MRS Communications 5 (2015) 265–275. https://doi.org/10.1557/mrc.2015.26.

[9] S. Sradhasagar, O. Subhasish Khuntia, S. Biswal, S. Purohit, A. Roy, Machine learning-aided discovery of bismuth-based transition metal oxide double perovskites for solar cell applications, Solar Energy 267 (2024) 112209. https://doi.org/10.1016/j.solener.2023.112209.

[10] R. Nechache, C. Harnagea, L.-P. Carignan, D. Ménard, A. Pignolet, Epitaxial Bi2FeCrO6 multiferroic thin films, Philosophical Magazine Letters 87 (2007) 231–240. https://doi.org/10.1080/09500830601153402.

[11] R. Nechache, C. Harnagea, L.-P. Carignan, O. Gautreau, L. Pintilie, M.P. Singh, D. Ménard, P. Fournier, M. Alexe, A. Pignolet, Epitaxial thin films of the multiferroic double perovskite Bi2FeCrO6 grown on (100)-oriented SrTiO3 substrates: Growth, characterization, and optimization, Journal of Applied Physics 105 (2009) 061621. https://doi.org/10.1063/1.3073826.

[12] R. Nechache, C. Harnagea, S. Licoccia, E. Traversa, A. Ruediger, A. Pignolet, F. Rosei, Photovoltaic properties of Bi2FeCrO6 epitaxial thin films, Applied Physics Letters 98 (2011) 202902. https://doi.org/10.1063/1.3590270.

[13] N.A. Mahammedi, A. Benameur, H. Gueffaf, B. Merabet, O.M. Ozkendir, S.-I. Sato, Investigating a Pb-free n-i-p perovskite solar cell with BFCO absorber using SCAPS-1D, Optik 302 (2024) 171659. https://doi.org/10.1016/j.ijleo.2024.171659.

[14] W. Huang, C. Harnagea, X. Tong, D. Benetti, S. Sun, M. Chaker, F. Rosei, R. Nechache, Epitaxial Bi2FeCrO6 Multiferroic Thin-Film Photoanodes with Ultrathin p-Type NiO Layers for Improved Solar Water Oxidation, ACS Applied Materials & Interfaces 11 (2019) 13185–13193. https://doi.org/10.1021/acsami.8b20998.

[15] S. Li, B. AlOtaibi, W. Huang, Z. Mi, N. Serpone, R. Nechache, F. Rosei, Epitaxial Bi2FeCrO6 Multiferroic Thin Film as a New Visible Light Absorbing Photocathode Material, Small 11 (2015) 4018–4026. https://doi.org/doi.org/10.1002/smll.201403206.

[16] L. Wei, J. Guo, L. Guan, B. Liu, First-principles investigation of bandgap tailoring in tetragonal Bi2FeCrO6 by magnetic ordering and B-site-cation ordering, Functional Materials Letters 13 (2019) 1950092. https://doi.org/10.1142/S1793604719500929.

[17] R. Nechache, C. Harnagea, S. Li, L. Cardenas, W. Huang, J. Chakrabartty, F. Rosei, Bandgap tuning of multiferroic oxide solar cells, Nature Photonics 9 (2014) 61–67. https://doi.org/10.1038/nphoton.2014.255.

[18] W. Huang, J. Chakrabartty, C. Harnagea, D. Gedamu, I. Ka, M. Chaker, F. Rosei, R. Nechache, Highly Sensitive Switchable Heterojunction Photodiode Based on Epitaxial Bi2FeCrO6 Multiferroic Thin Films, ACS Applied Materials & Interfaces 10 (2018) 12790–12797. https://doi.org/10.1021/acsami.8b00459.

[19] B. Merabet, H. Alamri, M. Djermouni, A. Zaoui, S. Kacimi, A. Boukortt, M. Bejar, Optimal Bandgap of Double Perovskite La-Substituted Bi2FeCrO6 for Solar Cells: an ab initio GGA+U Study, Chinese Physics Letters 34 (2017) 016101. https://doi.org/10.1088/0256-307X/34/1/016101.

[20] A. Raj, M. Kumar, A. Kumar, K. Singh, S. Sharma, R.C. Singh, M.S. Pawar, M.Z.A. Yahya, A. Anshul, Comparative analysis of 'La' modified BiFeO3-based perovskite solar cell devices for high conversion efficiency, Ceramics International 49 (2023) 1317–1327. https://doi.org/https://doi.org/10.1016/j.ceramint.2022.09.112.

[21] C. Senthilkumar, F.W. Shashikanth, Room temperature weak ferromagnetism in new Bi2MnCrO6 synthesized by gel combustion method, Applied Physics A 128 (2022) 244. https://doi.org/10.1007/s00339-022-05400-8.

[22] G. Kumar, B.K. Ravidas, S. Bhattarai, M.K. Roy, D.P. Samajdar, Exploration of the photovoltaic properties of oxide-based double perovskite Bi2FeCrO6 using an amalgamation of DFT with spin–orbit coupling effect and SCAPS-1D simulation approaches, New J. Chem. 47 (2023) 18640–18658. https://doi.org/10.1039/D3NJ02841K.

[23] S.K. Gupta, A. Kumar, S. Barthwal, Sadanand, N. Garg, C.K. Gupta, V. Yadav, S. Sharma, D.C. Tripathi, S. Kumar, SCAPS-1D study on the design and performance optimization of Sr3NCl3 solar cell: Assessing the significance of copper oxide (Cu2O) and copper(I) thiocyanate (CuSCN) as hole transport layers, Journal of Physics and Chemistry of Solids 208 (2026) 113170. https://doi.org/10.1016/j.jpcs.2025.113170.

[24] L.A. Lotfy, M. Abdelfatah, S.W. Sharshir, A.A. El-Naggar, W. Ismail, A. El-Shaer, Numerical simulation and optimization of FTO/TiO2/CZTS/CuO/Au solar cell using SCAPS-1D, Scientific Reports 15 (2025) 28022. https://doi.org/10.1038/s41598-025-12999-0.

[25] A. Saidarsan, S. Guruprasad, A. Malik, P. Basumatary, D.S. Ghosh, A critical review of unrealistic results in SCAPS-1D simulations: Causes, practical solutions and roadmap ahead, Solar Energy Materials and Solar Cells 279 (2025) 113230. https://doi.org/10.1016/j.solmat.2024.113230.

[26] N. Bhardwaj, A. Gaur, K. Yadav, Effect of doping on optical properties in BiMn1−x(TE)xO3 (where x = 0.0, 0.1 and TE = Cr, Fe, Co, Zn) nanoparticles synthesized by microwave and

sol-gel methods, Applied Physics A 123 (2017) 429. https://doi.org/10.1007/s00339-017-1042-y.

[27] F. Bai, L. Shi, H. Zhang, Z. Zhong, W. Wang, D. He, Multiferroic properties of La-doped Bi2FeCrO6 prepared by high-pressure synthesis, Journal of Applied Physics 111 (2012) 07C702. https://doi.org/10.1063/1.3670576.

[28] W. Xu, J. Sun, X. Xu, G. Yuan, Y. Zhang, J. Liu, Z. Liu, Reproducible resistive switching in the super-thin Bi2FeCrO6 epitaxial film with SrRuO3 bottom electrode, Applied Physics Letters 109 (2016) 152903. https://doi.org/10.1063/1.4964603.

[29] P. Mandal, A. Iyo, Y. Tanaka, A. Sundaresan, C.N.R. Rao, Structure, magnetism and giant dielectric constant of BiCr0.5Mn0.5O3 synthesized at high pressures, Journal of Materials Chemistry 20 (2010) 1646–1650. https://doi.org/10.1039/B914350P.

[30] A. Quattropani, D. Stoeffler, T. Fix, G. Schmerber, M. Lenertz, G. Versini, J.L. Rehspringer, A. Slaoui, A. Dinia, S. Colis, Band-Gap Tuning in Ferroelectric Bi2FeCrO6 Double Perovskite Thin Films, The Journal of Physical Chemistry C 122 (2018) 1070–1077. https://doi.org/10.1021/acs.jpcc.7b10622.

[31] R. Nechache, C. Harnagea, A. Pignolet, F. Normandin, T. Veres, L.-P. Carignan, D. Ménard, Growth, structure, and properties of epitaxial thin films of first-principles predicted multiferroic Bi2FeCrO6, Applied Physics Letters 89 (2006) 102902. https://doi.org/10.1063/1.2346258.

[32] L. Qiao, K.H.L. Zhang, M.E. Bowden, T. Varga, V. Shutthanandan, R. Colby, Y. Du, B. Kabius, P.V. Sushko, M.D. Biegalski, S.A. Chambers, The Impacts of Cation Stoichiometry and Substrate Surface Quality on Nucleation, Structure, Defect Formation, and Intermixing in Complex Oxide Heteroepitaxy–LaCrO3 on SrTiO3(001), Advanced Functional Materials 23 (2013) 2953–2963. https://doi.org/doi.org/10.1002/adfm.201202655.

[33] S. Yuan, Z. Mu, L. Lou, S. Zhao, D. Zhu, F. Wu, Broadband NIR-II phosphors with Cr4+ single activated centers based on special crystal structure for nondestructive analysis, Ceramics International 48 (2022) 26884–26893. https://doi.org/10.1016/j.ceramint.2022.05.391.

[34] H. Wu, Z. Pei, W. Xia, Y. Lu, K. Leng, X. Zhu, Structural, magnetic, dielectric and optical properties of double-perovskite Bi2FeCrO6 ceramics synthesized under high pressure,

Journal of Alloys and Compounds 819 (2020) 153007. https://doi.org/doi.org/10.1016/j.jallcom.2019.153007.

[35] G.A. Gomez-Iriarte, A. Pentón-Madrigal, L.A.S. De Oliveira, J.P. Sinnecker, XPS Study in BiFeO3 Surface Modified by Argon Etching, Materials 15 (2022) 4285. https://doi.org/10.3390/ma15124285.

[36] J.-W. Shi, C. Gao, C. Liu, Z. Fan, G. Gao, C. Niu, Porous MnOx for low-temperature NH3-SCR of NOx: the intrinsic relationship between surface physicochemical property and catalytic activity, Journal of Nanoparticle Research 19 (2017) 194. https://doi.org/10.1007/s11051-017-3887-6.

[37] C. Chen, X.-T. Wang, J.-H. Zhong, J. Liu, G.I.N. Waterhouse, Z.-Q. Liu, Epitaxially Grown Heterostructured SrMn3O6−x-SrMnO3 with High-Valence Mn3+/4+ for Improved Oxygen Reduction Catalysis, Angewandte Chemie International Edition 60 (2021) 22043–22050. https://doi.org/doi.org/10.1002/anie.202109207.

[38] M.T. Anderson, K.B. Greenwood, G.A. Taylor, K.R. Poeppelmeier, B-cation arrangements in double perovskites, Progress in Solid State Chemistry 22 (1993) 197–233. https://doi.org/10.1016/0079-6786(93)90004-B.

[39] Y. Li, H. Wu, W. Qi, X. Zhou, J. Li, J. Cheng, Y. Zhao, Y. Li, X. Zhang, Passivation of defects in perovskite solar cell: From a chemistry point of view, Nano Energy 77 (2020) 105237. https://doi.org/10.1016/j.nanoen.2020.105237.

[40] D.-Y. Son, S.-G. Kim, J.-Y. Seo, S.-H. Lee, H. Shin, D. Lee, N.-G. Park, Universal Approach toward Hysteresis-Free Perovskite Solar Cell via Defect Engineering, J. Am. Chem. Soc. 140 (2018) 1358–1364. https://doi.org/10.1021/jacs.7b10430.

[41] M. Abdi-Jalebi, M. Pazoki, B. Philippe, M.I. Dar, M. Alsari, A. Sadhanala, G. Divitini, R. Imani, S. Lilliu, J. Kullgren, H. Rensmo, M. Grätzel, R.H. Friend, Dedoping of Lead Halide Perovskites Incorporating Monovalent Cations, ACS Nano 12 (2018) 7301–7311. https://doi.org/10.1021/acsnano.8b03586.

[42] G. Ren, W. Han, Z. Li, C. Liu, L. Shen, W. Guo, Alkali metal ions passivation to decrease interface defects of perovskite solar cells, Solar Energy 193 (2019) 220–226. https://doi.org/10.1016/j.solener.2019.09.056.

[43] Y. Guo, F. Zhao, J. Tao, J. Jiang, J. Zhang, J. Yang, Z. Hu, J. Chu, Efficient and Hole-Transporting-Layer-Free CsPbI2Br Planar Heterojunction Perovskite Solar Cells through

Rubidium Passivation, ChemSusChem 12 (2019) 983–989. https://doi.org/10.1002/cssc.201802690.

[44] X. Liu, Y. Zhang, L. Shi, Z. Liu, J. Huang, J.S. Yun, Y. Zeng, A. Pu, K. Sun, Z. Hameiri, J.A. Stride, J. Seidel, M.A. Green, X. Hao, Exploring Inorganic Binary Alkaline Halide to Passivate Defects in Low-Temperature-Processed Planar-Structure Hybrid Perovskite Solar Cells, Advanced Energy Materials 8 (2018) 1800138. https://doi.org/10.1002/aenm.201800138.

[45] N.E. Grant, S.L. Pain, E. Khorani, R. Jefferies, A. Wratten, S. McNab, D. Walker, Y. Han, R. Beanland, R.S. Bonilla, J.D. Murphy, Activation of Al2O3 surface passivation of silicon: Separating bulk and surface effects, Applied Surface Science 645 (2024) 158786. https://doi.org/10.1016/j.apsusc.2023.158786.

[46] F.F. Targhi, Y.S. Jalili, F. Kanjouri, MAPbI3 and FAPbI3 perovskites as solar cells: Case study on structural, electrical and optical properties, Results in Physics 10 (2018) 616–627. https://doi.org/10.1016/j.rinp.2018.07.007.

[47] Ł. Haryński, A. Olejnik, K. Grochowska, K. Siuzdak, A facile method for Tauc exponent and corresponding electronic transitions determination in semiconductors directly from UV–Vis spectroscopy data, Optical Materials 127 (2022) 112205. https://doi.org/doi.org/10.1016/j.optmat.2022.112205.

[48] A. Quattropani, A.S. Makhort, M.V. Rastei, G. Versini, G. Schmerber, S. Barre, A. Dinia, A. Slaoui, J.-L. Rehspringer, T. Fix, S. Colis, B. Kundys, Tuning photovoltaic response in Bi2FeCrO6 films by ferroelectric poling, Nanoscale 10 (2018) 13761–13766. https://doi.org/10.1039/C8NR03137A.

[49] W. Huang, C. Harnagea, D. Benetti, M. Chaker, F. Rosei, R. Nechache, Multiferroic Bi2FeCrO6 based p–i–n heterojunction photovoltaic devices, Journal of Materials Chemistry A 5 (2017) 10355–10364. https://doi.org/10.1039/C7TA01604B.

[50] R. Nechache, L.-P. Carignan, L. Gunawan, C. Harnagea, G.A. Botton, D. Ménard, A. Pignolet, Epitaxial thin films of multiferroic Bi2FeCrO6 with B-site cationic order, Journal of Materials Research 22 (2007) 2102–2110. https://doi.org/10.1557/jmr.2007.0273.

[51] B. Nath, P.C. Ramamurthy, D.R. Mahapatra, G. Hegde, Effect of cuprous iodide passivation in perovskite solar cells, Journal of Materials Science: Materials in Electronics 33 (2022) 14457–14467. https://doi.org/10.1007/s10854-022-08368-6.

[52] T.D. Subha, R.T. Prabu, S. Parasuraman, A. Kumar, Role of Urbach energy in controlling voltage output of solar cells, Optical and Quantum Electronics 55 (2023) 794. https://doi.org/10.1007/s11082-023-05067-2.

[53] W. Li, K. Zhao, H. Zhou, W. Yu, J. Zhu, Z. Hu, J. Chu, Precursor solution temperature dependence of the optical constants, band gap and Urbach tail in organic–inorganic hybrid halide perovskite films, J. Phys. D: Appl. Phys. 52 (2019) 045103. https://doi.org/10.1088/1361-6463/aaec21.

[54] F. Urbach, The Long-Wavelength Edge of Photographic Sensitivity and of the Electronic Absorption of Solids, Physical Review 92 (1953) 1324. https://doi.org/10.1103/PhysRev.92.1324.

[55] J.D. Dow, D. Redfield, Toward a Unified Theory of Urbach's Rule and Exponential Absorption Edges, Physical Review B 5 (1972) 594–610. https://doi.org/10.1103/PhysRevB.5.594.

[56] G.D. Cody, T. Tiedje, B. Abeles, B. Brooks, Y. Goldstein, Disorder and the Optical-Absorption Edge of Hydrogenated Amorphous Silicon, Physical Review Letters 47 (1981) 1480–1483. https://doi.org/10.1103/PhysRevLett.47.1480.

[57] E. Arushanov, L. Kulyuk, O. Kulikova, V. Tezlevan, R.F. Ruiz, M. León, Optical study of monocrystalline $CuIn_4 Se_6$, J. Phys. D: Appl. Phys. 34 (2001) 3480–3484. https://doi.org/10.1088/0022-3727/34/24/309.

[58] M. Kumar, S.K. Pundir, D.V. Singh, M. Kumar, Effect on green energy conversion and stability with 'Er' modification in multiferroic based perovskite solar cell devices, Materials Today Communications 38 (2024) 107841. https://doi.org/10.1016/j.mtcomm.2023.107841.

[59] J. Li, H.A. Dewi, H. Wang, J. Zhao, N. Tiwari, N. Yantara, T. Malinauskas, V. Getautis, T.J. Savenije, N. Mathews, S. Mhaisalkar, A. Bruno, Co-Evaporated MAPbI3 with Graded Fermi Levels Enables Highly Performing, Scalable, and Flexible p-i-n Perovskite Solar Cells, Advanced Functional Materials 31 (2021) 2103252. https://doi.org/doi.org/10.1002/adfm.202103252.

[60] P. Murugan, T. Hu, X. Hu, Y. Chen, Current Development toward Commercialization of Metal-Halide Perovskite Photovoltaics, Advanced Optical Materials 9 (2021) 2100390. https://doi.org/doi.org/10.1002/adom.202100390.

[61] H.-Y. Tu, X. Qi, Growth of p/n-type BiFeO3 thin films for construction of a bilayer p–n junction for photodegradation of organic pollutants, Journal of Materials Chemistry A 12 (2024) 12752–12761. https://doi.org/10.1039/D4TA01615G.

[62] M.T. Neukom, S. Züfle, E. Knapp, M. Makha, R. Hany, B. Ruhstaller, Why perovskite solar cells with high efficiency show small IV-curve hysteresis, Solar Energy Materials and Solar Cells 169 (2017) 159–166. https://doi.org/doi.org/10.1016/j.solmat.2017.05.021.

[63] J.C. Wang, X.C. Ren, S.Q. Shi, C.W. Leung, P.K.L. Chan, Charge accumulation induced S-shape J–V curves in bilayer heterojunction organic solar cells, Organic Electronics 12 (2011) 880–885. https://doi.org/10.1016/j.orgel.2011.02.016.

[64] P. Arnou, C.S. Cooper, A.V. Malkov, J.W. Bowers, J.M. Walls, Solution-processed CuIn(S,Se)2 absorber layers for application in thin film solar cells, Thin Solid Films 582 (2015) 31–34. https://doi.org/doi.org/10.1016/j.tsf.2014.10.080.

[65] A. El Badraoui, T. Chargui, A. Elkhou, B. El Mokhtari, R. El Idrissi, L. El Amri, H. Ez-Zahraouy, N. Tahiri, Optimization of lead-free CsSnBr3-based perovskite solar cells via SCAPS-1D simulations and machine learning models, Journal of Physics and Chemistry of Solids 210 (2026) 113340. https://doi.org/10.1016/j.jpcs.2025.113340.

[66] V.H.D. Araújo, A.F. Nogueira, J.C. Tristão, L.J. dos Santos, Advances in lead-free perovskite solar cell design via SCAPS-1D simulations, RSC Sustainability 3 (2025) 4314–4335. https://doi.org/10.1039/D5SU00526D.